\documentclass[aps,prl,twocolumn,
showpacs,
floatfix,nofootinbib,
superscriptaddress]{revtex4}

\usepackage{epsfig}

\newcommand{\be}{\begin{equation}}
\newcommand{\ee}{\end{equation}}
\newcommand{\bear}{\begin{eqnarray}}
\newcommand{\eear}{\end{eqnarray}} \newcommand{\ba}{\begin{array}}
\newcommand{\ea}{\end{array}}

\newcommand{\gae}{\begin{array}{c}\,\sim\vspace{-1.7em}\\> 
\end{array}}

\def\beq{\begin{equation}}
\def\eeq#1{\label{#1}\end{equation}}
\def\eeqn{\end{equation}}
\def\eeq{\end{equation}}
\def\beqa{\begin{eqnarray}}
\def\eeqa#1{\label{#1}\end{eqnarray}}
\def\eeqan{\end{eqnarray}}

\def\to{\rightarrow}

\newcommand\iden{\leavevmode\hbox{\small1\normalsize\kern-.33em1}}

\def\W3{W_H^3}

\begin{document}

\title{A Signal for a Theory of Flavor at the Large
Hadron Collider}
\author{Priscila M. Aquino\footnote{E-mail: priscila@fma.if.usp.br}}
\author{
Gustavo Burdman\footnote{E-mail: burdman@fma.if.usp.br}} 
\author{Oscar~J.~P.~\'{E}boli\footnote{E-mail: eboli@fma.if.usp.br}}
\affiliation{Instituto de F\'isica, Universidade de S\~ao Paulo, 
S\~ao Paulo SP 05508-900, Brazil}
\vspace*{0.3cm}
%\date{\today}

%\vspace*{0.4in}

\begin{abstract}
%\vspace*{0.2in}
In Randall-Sundrum models with gauge bosons and fermions in the extra dimensional 
bulk, it is possible to build models of flavor by localizing the 
fermions in the extra dimension. Since the Higgs must be localized at or close to 
the TeV scale fixed point, heavier fermions must be localized close to this 
brane. The first Kaluza-Klein excitations of  the gauge bosons are also 
TeV-localized, so they have stronger couplings to heavier fermions leading to 
tree-level flavor-violating couplings. We investigate
the potential of the LHC to observe flavor violation in single top production
at very high invariant masses, in addition to the observation of the corresponding
top--anti-top resonance. We conclude that the LHC will be able to observe 
tree-level flavor violation in single top production, probing 
KK masses at least as large as $2$~TeV, as well as 
a very interesting  region of the parameters.

\end{abstract}
%}

\maketitle

\paragraph{Introduction}
It is generally believed that extensions of the 
standard model must address the stability of the weak scale 
with new physics at energies not far above the TeV scale. 
The origin of fermion masses, on the other hand, could in principle reside
at much higher scales. Nonetheless, it is tempting to consider the possibility
that the origin of at least some of the fermion masses is 
related to TeV scale physics.
This is particularly true when considering the origin of the top
quark mass, which is of the order of the weak scale. The correspondingly
large Yukawa coupling suggests the presence of a strongly coupled sector
associated with the third generation quarks.

Recently, Randall and Sundrum proposed the use 
of a non-factorizable geometry in five dimensions~\cite{rs1} as a solution of the 
hierarchy problem. 
The extra dimension is compactified
on an orbifold $S_1/Z_2$ of radius $r$ so that the bulk is a slice of ${\rm AdS}_5$
space between two four-dimensional boundaries.
The metric depends on the five dimensional coordinate $y$ and is given by~\cite{rs1} 
\begin{equation}
ds^2 = e^{-2\sigma(y)} \eta_{\mu\nu} dx^\mu dx^\nu - dy^2~,
\label{metric}
\end{equation} 
where $x^\mu$ are the four dimensional coordinates, $\sigma(y) = k |y|$, with 
$k\sim M_P$ characterizing the curvature scale.  This metric generates
two effective scales: $M_P$ and $M_P e^{-k\pi r}$. In this way, values of 
$r$ not much larger than the Planck length ($kr\simeq (11-12)$)
can be used in order to generate
a scale $\Lambda_r\simeq M_Pe^{-k\pi r}\simeq~{\rm O(TeV)}$ on one of the boundaries.

In the original RS scenario, only gravity was allowed to propagate in the bulk, 
with the Standard Model (SM) fields confined to one of the boundaries. 
The inclusion of matter and gauge fields in the bulk has been extensively treated in the
literature~\cite{bulk1,chang,gn,gp,huber1,dhr1,dhr2}. 
The Higgs field must be localized on or around the TeV brane in order to generate the 
weak scale. 
As it was recognized in Ref.~\cite{gp}, it is possible to generate
the fermion mass hierarchy from $O(1)$ flavor breaking in the bulk masses of fermions. 
Since bulk fermion masses result in the localization of fermion zero-modes, 
lighter fermions 
should be localized toward the Planck brane, where their wave-functions have an 
exponentially
suppressed overlap with the TeV-localized Higgs, whereas fermions with order one
Yukawa couplings should be localized toward the TeV brane. 

Since the lightest KK excitations of gauge
bosons are localized toward the TeV brane, they tend to be strongly coupled to 
zero-mode fermions localized there. Thus, the flavor-breaking fermion localization 
leads to flavor-violating interactions of the KK gauge bosons, particularly with third
generation quarks. For instance, the first KK excitation of the gluon, will have 
flavor-violating  neutral couplings such as $G_\mu^a(t\gamma^\mu T^a\bar q)$, where $q=u,c$.
Consequences of such effects have been considered in Ref.~\cite{fvinrs}. 
Here, we will study the potential of the LHC to directly observe the 
flavor-violating couplings through the anomalous production of single top quarks
at very high invariant masses.

The action for fermion fields in the bulk contains a bulk mass term 
as in~\cite{chang,gn}
\begin{equation}
S_f = \int d^4x~dy~ \sqrt{g} \left\{ 
\cdots
- {\rm sgn}(y) M_f \bar\Psi\Psi\right\}~,
\label{sfermions} 
\end{equation}
where the dots stand for the fermion kinetic terms.
The bulk mass term $M_f$ in eqn.~(\ref{sfermions}) is expected to be of order
$k\simeq M_P$. 
Although the fermion field $\Psi$ is non-chiral, we can still define 
$\Psi_{L,R}\equiv \frac{1}{2}(1\mp\gamma_5)\Psi$. The KK decomposition can be written as
\begin{equation}
\Psi_{L,R}(x,y) = \frac{1}{\sqrt{2\pi r}}\,\sum_{n=0}\,\psi_n^{L,R}(x) e^{2\sigma} 
f_n^{L,R}(y)~,
\label{kkfermion}
\end{equation}
where $\psi_n^{L,R}(x)$ corresponds to the $nth$ KK fermion excitation and is 
a chiral four-dimensional field. 
The zero mode wave functions are 
\begin{equation}
f_0^{R,L}(y) = \sqrt{\frac{2k\pi r\,(1\pm 2c_{R,L})}{e^{k\pi r(1\pm2c_{R,L})}-1}}\;
e^{\pm c_{R,L} \,k\,y}~,
\label{zeromode} 
\end{equation}
with $c_{R,L}\equiv M_f/k$ parametrizing the bulk fermion mass in units of the 
inverse AdS radius $k$, and naturally of $O(1)$. 
The $Z_2$ orbifold projection is used so that only 
one of these is actually allowed, either a left-handed or a right-handed zero mode.

Fermion bulk mass parameters $c_L^i$ and $c_R^i$ are constrained 
by the requirement that we obtain the correct fermion masses and 
mixings with 5D Yukawa couplings of order one. 
For instance, if for simplicity we assumed that 
the Higgs is localized on the TeV brane, 
the Yukawa couplings of bulk fermions can be written as 
\begin{equation}
S_Y = \int d^4x \,dy \,\sqrt{-g}\,\frac{\lambda_{ij}^{5D}}{2\,M_5} \,\bar{\Psi}_i(x,y) 
\delta(y-\pi r) H(x)
\Psi_j(x,y)~,
\label{yuka5d}
\end{equation}
where $\lambda_{ij}^{5D}$ is a dimensionless parameter and $M_5$ is the fundamental 
scale or cutoff
of the theory. In order for this model to be a natural 
solution to the fermion mass hierarchy problem, we 
need $|\lambda_{ij}^{5D}| \sim O(1)$. 
The resulting zero-mode fermion mass matrices have then the form
\be
M_{ij} = \frac{\lambda_{ij}^{5D}\,v}{2\pi r\,M_5}\,f_{0i}^L(\pi r)\,f_{0j}^R(\pi r)~,
\label{mij}
\ee
where $v\simeq 174~$GeV is the Higgs VEV.
Thus we can see that in order to get small masses, light fermions should be localized towards the Planck brane. 
This results in  
their couplings to KK gauge bosons being mostly universal leading to
no observable FCNC effects. On the other hand, 
third-generation quarks are required to be close to the TeV brane so as to generate 
a large enough top mass. 
This posses a potential problem: both the right handed top quark and 
the left-handed doublet $Q^3$, should be significantly localized 
towards the ``Higgs'' brane, meaning that they can have
rather strong couplings to the first KK excitations of gauge bosons.
This induced flavor violation of KK gauge bosons with $b_L$ (we assume $
b_R$ localized on the 
Planck brane) is, in principle, constrained by the precise measurement of the 
$Z\to b\bar b$ interactions at the $Z$-pole. Depending on the model, this leads
to a bound on $c_L^3$~\cite{agashe2}, as well as to a typical KK gauge boson mass bound of 
$\gae 3~$TeV. 
Also electroweak precision constraints require that the gauge symmetry in the bulk be 
enlarged to 
$SU(3)_c\times~SU(2)_L\times~SU(2)_R\times U(1)_{B-L}$. 
More recently, and in the context of these very same models, it has been argued that it is 
possible to protect the $Z\to b\bar b$ coupling from large corrections by new 
symmetries~\cite{adp}, 
resulting in potentially lower KK masses. 
Also, in models without a Higgs~\cite{higgsless} or with a bulk Higgs~\cite{gaugephobic}, 
a KK scale as low as $1$~TeV can be readily accommodated. 
For the purpose of this work, we will be only concerned with the strong interactions, 
and not with 
the details of the electroweak sector. Thus, to be general enough, 
we will consider KK gauge boson 
masses as low as $1$~TeV.

We will consider the range of parameters defined by $c_L^3= [0.3,0.4]$ and 
$c_R^t=[-0.4,0.1]$.
This is chosen such that although the correct top mass is obtained, 
no contradictions with electroweak measurements are present, particularly with the $Z\to b\bar b$ 
rate.
We will only take the  combinations of parameters in these ranges that satisfy 
the requirement imposed by the quark masses and mixings.

%\begin{figure}[t]
%
%\centering
%
%\epsfig{file=singlet.eps,width=3cm,height=2.5cm,angle=0}
%
%\caption{The flavor violating vertex of the $G_\mu^{(1)}$, the first KK excitation 
%of the gluon. The circle corresponds to the action of $U_{L,R}^{(tq)}$, in addition 
%of the coupling of the $t_{L,R}$ to the KK gluon.
%}
%
%\label{singlet}
%
%\end{figure}

\paragraph{Couplings}
We will assume that the only non-universal couplings are those of 
the $t_R$, $t_L$ and $b_L$ with the KK gauge bosons. All other fermions, 
including the right-handed b quark are assumed to be Planck brane localized. 
The non-universality of the KK gauge boson couplings leads to tree-level 
flavor violation. The diagonalization of the quark mass matrix requires 
a change of basis for the quarks fields. In the SM, this rotation leads 
to the CKM matrix in the charged current, but the universality of the 
gauge interactions results in the Glashow-Ilyopoulos-Maiani (GIM) mechanism 
in the neutral currents. However, since the KK excitations of the gauge bosons are 
non-universal, tree-level GIM-violating couplings will appear in the physical 
quark basis. 

We will consider the dominant non-universal effect as coming from the 
couplings of $t_R$, $t_L$ and $b_L$ to the first KK excitation of the 
gluon: $g_{t_R}$, $g_{t_L}$ and $g_{b_L}$ respectively. 
The $SU(2)_L$ bulk symmetry implies $g_{t_L}=g_{b_L}$. 
In order to obtain these effective 4D couplings we need to 
compute
\be
g_5\int d^4x\int dy \,\sqrt{g} \,\bar\Psi(x,y) \gamma^\mu T^a G^a_\mu(x,y) \Psi(x,y)
\label{g4d}
\ee
where $g_5$ is the 5D $SU(3)_c$ coupling, and 
the KK expansion of the gluon $G^a_\mu(x,y)$ and its wave-functions in the 
extra dimension can be found in \cite{bulk1}. The couplings of eqn.~(\ref{g4d}) 
depend on the fermion bulk masses. For the range given above for $c_L^3$ and 
$c_R^t$ we obtain $g_{t_L} = g_{b_L} = [1.0,2.8]\,g_s$ and $g_{t_R} = [1.5,5]\, g_s$, 
where $g_s$ is the usual 4D $SU(3)_c$ coupling. 
The universally coupled light quarks as well as the right-handed b quark have 
$g_L^q = g_R^q = g_R^b \simeq -0.2\, g_s$. 

We are now in a position to compute the effects of all the {\em flavor-conserving }
interactions of the $G_\mu^{a(1)}(x)$, the first KK excitation of the gluon. 
We can also compute the width of the KK gluon. 
This can be written as 
\be
\Gamma \simeq\frac{\alpha_s}{12}\, M_G \left(9\,{\tilde g_q^2} + 
2\, {\tilde g_{t_L}^2} 
+ {\tilde g_{t_R}^2}\right)
\label{width}
\ee
where we defined the $\tilde g_q$'s as the couplings in units of the strong coupling 
$g_s$, and we neglected the flavor-violating contributions. 
Thus, the range of couplings defined above will result in a minimum and a maximum
width for the KK gluon. These are $\Gamma_{\rm min.}\simeq 0.04\,M_G$ and 
$\Gamma_{\rm max.} \simeq 0.35\,M_G$. Then, we see that the range of values
for the couplings allow for rather narrow or rather broad resonances, two very different
scenarios from the point of view of the phenomenology. 

We can also estimate now the size of the flavor changing couplings.
These arise as a consequence of the rotation to the quark mass eigen-basis. Then, if the 
CKM matrix is generated correctly, in this physical basis the left-handed up-type quarks
couple to the KK gluon through the following currents: 
$U_L^{tt}\,(\bar t_L T^a \gamma_\mu t_L)$, 
$U_L^{tc}\,(\bar t_L T^a \gamma_\mu c_L)$ and $U_L^{tu}\,(\bar t_L T^a \gamma_\mu u_L)$. 
Similarly, the right-handed up-type quarks couple through 
$U_R^{tt}\,(\bar t_R T^a\gamma_\mu t_R)$, 
$U_R^{tc}\,(\bar t_R T^a \gamma_\mu c_R)$ and $U_R^{tu}\,(\bar t_R T^a\gamma_\mu u_R)$.
Here, $U_L$ and $U_R$ are the left-handed and right-handed up-type quark rotation 
matrices responsible for the diagonalization of the Yukawa couplings of the 
up-type quarks. 
These  are not observable in the SM, but here they 
govern the flavor-violating amplitudes. In principle, we have little knowledge of the
entries of these matrices since there are many possible choices for textures
that would result in the correct quark masses and mixings.  
However, given that we know that $V_{\rm CKM} = U_L^\dagger D_L$, we can assume that 
$U_L\simeq \sqrt{V_{\rm CKM}}$, and similarly for 
$D_L$. This leads to $U_L^{tc} \simeq V_{cb}\simeq 0.04$ 
and $U_L^{tu} \simeq V_{ub}\simeq 0.004$. 
On the  other hand, we have no bias from the SM on the entries of $U_R$. 
Constraints on its elements can be imposed by asking for the 
correct up-quark masses after diagonalization.  
However, values of $U_R^{tc}\simeq O(1)$ are still possible. 
For instance, if most of the charm quark mass comes from 
the mixing with the third generation up quark, then $U_R^{tc}=1$ is possible~\cite{usagain}. 
This is even more so if $\lambda_{23}^{5D} < \lambda_{33}^{5D}$, while both still of $O(1)$.
In what follows we will consider $U_R^{tc}$ and $U_R^{tu}$ as free parameters.
Since we will assume no separation of charm from light jets, we 
define $U_R^{tq}\equiv \sqrt{(U_L^{tc})^2 + (U_L^{tu})^2}$, and 
study the sensitivity of the LHC to this parameter for a given KK gluon mass.

\paragraph{Signal and Backgrounds}
These flavor-violating interactions could be directly 
observed by the s-channel production of the first KK excitation of the gluon
with its subsequent decay to a top and a charm or up quark. For instance, at the 
LHC we could have the reaction 
\be
p p\to G_\mu^{a(1)} \to t q~,
\label{singletop}
\ee
with $q=u,c$. 
Thus, the Randall-Sundrum scenario with bulk matter predicts anomalous 
single top production at a very high invariant mass, which is determined by the 
mass of the KK gluon. 
In what follows, we study in detail the potential of the LHC for observing
this signal, with a top quark and a jet with very high invariant mass.

We will present results
for the sets of values for $c_L^3$ and $c_R^t$ that result in the minimum 
width for the KK gluon, as estimated above. 

In order to reduce the backgrounds we 
considered only the semi-leptonic decays of the top quarks
%\[
$   p p  \to t \bar{q}\; (\bar{t} q) \to b \ell^+ \nu_\ell \bar{q} \;
(\bar{b} \ell^- \bar\nu_\ell q) $, 
%\;\; ,
%\]
where $\ell =e$ or $\mu$, and $q=u,c$.  
Therefore, this signal exhibits one b jet, one light
jet, a charged lepton and missing transverse energy.
There are many SM backgrounds for this process, namely: 

\begin{itemize}

\item $ p p \to t \bar{t} \to b \ell^+ \nu_\ell \bar{b} \ell^- \bar\nu_\ell$
  when one of the b jets is mistagged and one of the charged leptons is either
  lost in the beam pipe or embedded in one of the jets. It is interesting to
  notice that the flavor-conserving signal for the first KK state of the gluon 
  decaying to $t \bar{t}$ pairs  also contributes to this
  background. 

\item $pp \to W^\pm jj \to \ell^\pm \nu jj$ where one of the light jets is
tagged as a b jet. 

\item $pp \to W^\pm b \bar{b} \to \ell^\pm \nu b \bar{b}$ where one of the b
  jets is mistagged.

\item $pp \to W^{*\pm} \to t \bar{b} + \bar{t} b \to b \bar{b} \ell^\pm \nu $
where one of the b jet is mistagged.

\item $pp \to t b j \to b \bar{b} j \ell^\pm \nu$ where one of the jets is
  lost and just one jet is tagged as a b jet. Here in the $W$--gluon fusion we
  considered only the diagrams that are not higher order corrections to the
  single production of tops via $W^*$.

\end{itemize}
We simulate the signal and backgrounds with the package MADEVENT~\cite{madevent}.
We initially imposed the following jet and lepton acceptance cuts
\begin{eqnarray}
  && p_T^j > 20 \hbox{ GeV} \;\;\;\;\;\;\; , \;\;\;\; | y_j | < 2.5 \;\; ,
  \nonumber
  \\
  && p_{T}^\ell \geq 20 \; {\rm GeV} 
  \;\;\;\;\;\;\; , \;\;\;\;
  |y_\ell|\leq 2.5
\label{cuts1}
\\
&& \Delta R_{\ell j}\geq 0.63
  \;\;\;\;\;\;\; , \;\;\;\;
\Delta R_{\ell \ell}\geq 0.63 \;\; ,\nonumber 
\end{eqnarray}
where $j$ can be either a light or a b jet.

In order to further reduce the background we imposed the following additional cuts

\begin{enumerate}

\item We require that the invariant mass of the system formed by the lepton, the b
  tagged jet and the light jet be within a window 
\begin{equation}
  M_{G^{(1)}} - \Delta \le M_{bj\ell} \le M_{G^{(1)}} + \Delta
\label{cuts2}
\end{equation}
 around the first KK excitation
    of the gluon mass. This cut ensures that the selected events have large invariant masses, 
as required by the large mass of the s-channel object being exchanged.
The values of $\Delta$ used in this study are presented in Table~\ref{t1}.

\item We require the transverse momentum of the light jet to be larger than
$p_{\rm cut}$, i.e.,
\begin{equation}
    p_{j~light} \ge p_{\rm cut}
\label{cuts3}
\end{equation}
Since the light jet in the signal recoils against the top forming with it a large invariant mass, 
it tends to be harder than the jets occurring in the background.
We present in Table~\ref{t1} the values for $p_{\rm cut}$ used in our analysis.

\item We also require that the invariant mass of the charged lepton 
and the b tagged jet
be {\em smaller} than 250 GeV:
\begin{equation}
  M_{b\ell} \le 250 \hbox{ GeV}~.
\label{cuts4}
\end{equation}
This requirement is always passed by the signal, but eliminates a sizable fraction of the 
$Wjj$ background. It substitutes for the full top reconstruction when the neutrino momentum 
is inferred, which we are not using here.

\end{enumerate}

\begin{table}
\begin{center}
\begin{tabular}{|c|c|c|}
\hline
$M_{G^{(1)}}$ (TeV)      &    $\Delta$ (GeV)   & $p_{\rm cut}$ (GeV)
\\
\hline
1     &     120   &  350
\\
\hline
2     &     250   &  650
\\
\hline

\end{tabular}
\end{center}
\caption{Cuts used in our analysis. See text for details.}
\label{t1}
\end{table}

We present in Table~\ref{t2} the cross sections for signal and backgrounds
for $M_{G^{(1)}}=1~$TeV. The main sources of backgrounds are $Wjj$ and $t\bar t$ 
production. The latter includes  the 
{\em KK gluon flavor conserving signal}. 
Thus, this background partly scales with $M_G$ the same way the signal does. 
The signal is obtained for $U_R^{tq} = 1$ and neglecting the contributions from 
left-handed final states, corresponding to $U_L^{tq}=0$. 
The latter approximation is justified 
since we assume the left-handed quark rotations to be proportional to
the CKM matrix, a conservative assumption.
Regarding the choice of bulk masses, we fix these so as to obtain the minimum width which, as 
mentioned above, can be as small as $\Gamma_G\simeq 0.04 M_G$. 
Table~\ref{t3} shows the results for $M_G = 2~$TeV under similar assumptions. 

In order to evaluate the reach of the LHC, we will require a significance
of $5\,\sigma$ for the signal over the 
background. For a given KK gluon mass and accumulated luminosity, 
this can be translated into a reach in the flavor-violating
parameter $U_R^{tq}$ defined above. This is shown in Table~\ref{reach}. 
We see that the LHC will  be sensitive to tree-level flavor violation 
for KK gluon masses of up to at least $2~$TeV, probing a very interesting region of values for 
$U_R^{tq}$. The reach can be somewhat better if we allow for the reconstruction of the 
momentum of the neutrino coming from the $W$ decay, which typically reduces the $Wjj$ 
background more drastically.

\begin{table}
\begin{center}
\begin{tabular}{|c|c|c|c|}
\hline
Process       
& $\sigma ~-$  (\ref{cuts2}) & $\sigma ~-$ (\ref{cuts3})
& $\sigma ~-$ (\ref{cuts4})
\\
\hline
$pp\to tj$    &  148 fb   & 103 fb & 103 fb
\\
\hline
$pp \to Wjj$                    &  243 fb   &  42.0 fb & 21.0 fb
\\
\hline
$pp \to Wbb$                  &  11.1 fb  & 4.07 fb & 3.19 fb
\\
\hline
$pp \to tb$                       &  1.53 fb  & 0.70 fb & 0.61 fb
\\
\hline
$pp \to t \bar{t}$               &  44.4 fb  & 15.1 fb & 14.2 fb
\\
\hline
$Wg$ fusion                  &  32.0 fb  & 5.23 fb & 5.23 fb
\\
\hline
\end{tabular}
\end{center}
\caption{Signal and background cross sections for a KK gluon of $M_{G^{(1)}} = 1$~TeV, 
after the successive 
application of the cuts defined in (\ref{cuts2}), (\ref{cuts3}) and (\ref{cuts4}). 
  Efficiencies and b tagging probabilities are already included. Here we 
used $U_R^{tq}=1$.}
\label{t2}
\end{table}

\paragraph{Conclusions}
We have performed the first study of flavor violation at the LHC  mediated by a high invariant mass
object. Unlike rare top decays or low energy processes, this signal can only be a consequence 
of the tree-level flavor violation characteristic in these models.
These phenomenon is expected in theories with one extra dimension with Anti--de Sitter 
curvature 
in the bulk, where gauge and matter fields are allowed to propagate in the bulk, and  
{\em is an essential manifestation of a theory of flavor}.
Its observation would constitute a primary test of this aspect of bulk Randall-Sundrum models. 
We conclude that the LHC has the potential to observe this phenomenon
up to KK gluon masses of at least $M_{G^{(1)}} = 2~$TeV for interesting 
values of the parameter
$U_R^{tq}$ (see Table~\ref{reach}).  In this study, we used narrow resonances. A more thorough study of 
the LHC reach for broader resonances is needed. Among other aspects for further study are  
the impact on the background of using a full top reconstruction, and of radiative corrections and the choice 
of renormalization scale. We leave these more detailed studies for future work~\cite{usagain}.

\begin{table}
\begin{center}
\vspace*{0.2in}
\begin{tabular}{|c|c|c|c|}
\hline
Process     
& $\sigma ~-$  (\ref{cuts2}) & $\sigma ~-$ (\ref{cuts3})
& $\sigma ~-$ (\ref{cuts4})
\\
\hline
$pp\to tj$    &  5.10 fb   & 2.18  fb& 2.18 fb
\\
\hline
$pp \to Wjj$   & 25.4 fb   & 3.79 fb & 0.95 fb
\\
\hline
$pp \to Wbb$   & 0.97 fb & 0.45 fb & 0.06  fb
\\
\hline
$pp \to tb$      & 0.04 fb  & 0.02 fb & 0.02 fb
\\
\hline
$pp \to t \bar{t}$   & 1.60 fb  & 0.29 fb & 0.24 fb
\\
\hline
$Wg$ fusion          & 1.20 fb  & 0.10 fb & 0.10 fb
\\
\hline
\end{tabular}
\end{center}
\caption{Signal and background cross sections for a KK gluon of $M_{G^{(1)}} = 2$~TeV. 
  Efficiencies and b tagging probabilities are already included.
Here we used $U_R^{tq}=1$}
\label{t3}
\end{table}

\begin{table}
\begin{center}
\begin{tabular}{|c|c|c|c|}
\hline
$M_G~$ [TeV] & $30 fb^{-1}$ & $100 fb^{-1}$ & $300 fb^{-1}$
\\
\hline
1            & 0.24         & 0.18          & 0.14
\\
\hline
2            & 0.65         & 0.50          & 0.36
\\
\hline
\end{tabular}
\end{center}
\caption{Reach in $U_R^{tq}$ for various integrated luminosities. }
\label{reach}
\end{table}

\paragraph{Acknowledgments } 
The authors acknowledge the support of the Brazilian  National Counsel
for Technological and Scientific Development (CNPq).
G.~B. and O.~E. also acknowledge the support of the State of S\~{a}o Paulo
Research Foundation (FAPESP).

\end{document}